\documentclass[twocolumn,prappl]{revtex4}

  \usepackage[dvips]{graphicx}

\usepackage{dcolumn}
\usepackage{amsmath}
\usepackage{latexsym}

\begin{document}

\title[Short Title]{Josephson traveling-wave parametric amplifier with three-wave mixing}
\author{A. B. Zorin}

\affiliation{Physikalisch-Technische Bundesanstalt, Bundesallee
100, 38116 Braunschweig, Germany}
\affiliation{Department of Physics, Lomonosov Moscow State University, Moscow, 119991, Russia}

\date{February 26, 2016, revised August 2, 2016}

\begin{abstract}
We develop a concept of the traveling-wave Josephson parametric amplifier
exploiting quadratic nonlinearity of a serial array of one-junction SQUIDs
embedded in a superconducting transmission line.
The external magnetic flux applied to the SQUIDs makes it possible
to efficiently control the shape of their current-phase relation and, hence,
the balance between quadratic and cubic (Kerr-like) nonlinearities.
This property allows us to operate in the favorable three-wave-mixing mode with minimal phase mismatch,
an exponential dependence of the power gain on number of sections $N$, a large bandwidth,
a high dynamic range, and substantially separated
signal ($\omega_s$) and pump ($\omega_p$) frequencies obeying relation
$\omega_s+\omega_i = \omega_p$, where $\omega_i$ is the idler frequency.
An estimation of the amplifier characteristics with typical experimental
parameters, a pump frequency of $12$~GHz, and $N = 300$ yields
a flat gain of 20~dB in the bandwidth of $5.6$~GHz.


\verb  PACS numbers: 84.30.Le, 85.25.Cp, 05.45.-a, 74.81.Fa

\end{abstract}
\maketitle

\section{\label{sec:Intro}Introduction}

These days, the Josephson parametric amplifiers (JPAs) \cite{Yurke1989} have practically achieved
a quantum-limited performance \cite{Movshovich1990,Castellanos-Beltran2007,Mutus2014} and
are considered to be the most advanced tools available
for fine experiments in the field of quantum measurements \cite{Hatridgel2011,Teufel2011}
and quantum-information technologies \cite{Abdo2011,Eichler2012,Abdo2013,Krantz2015}.
Recently, owing to impact of the kinetic-inductance traveling-wave parametric
amplifier \cite{Eom2012}, the Josephson traveling-wave
parametric amplifiers (JTWPAs) enabling larger gain per unit length with lesser pump power have
been in the particular focus of several research groups
\cite{Yaakobi2013,OBrien2014,White2015,Bell-Samolov2015,Macklin2015}. Moreover,
these promising devices have already demonstrated the performance with noise approaching the quantum limit
(see the works of White et al. \cite{White2015} and Macklin et al. \cite{Macklin2015}).
In contrast to conventional JPAs including Josephson junctions (JJs) embedded
in superconducting cavities and, therefore, suffering from inherent gain-bandwidth trade-off,
JTWPAs are designed as microwave transmission lines enabling the mixing
of propagating microwaves and, therefore, free of the bandwidth limitation and allowing
higher dynamic range. These properties are required particularly for the
frequency-multiplexing readout of quantum objects \cite{Jeffrey2014}.

Similar to the concept of parametric amplification in nonlinear optical fibers \cite{Agrawal},
JTWPAs analyzed \cite{Yaakobi2013,OBrien2014,Bell-Samolov2015} and
accomplished \cite{White2015,Macklin2015} to date were based on the Kerr
nonlinearity, i.e. on the dependence of the refractive index $n$
on intensity of the wave $\propto|\textbf{\textit{E}}|^2$.
In superconducting circuits, this nonlinearity is due to the
dependence of the Josephson inductance (equivalent to refractive
index $n$ in optics) on the square of current $I^2$,
viz. $L_J(I) \approx \Phi_0 (1 + \tilde{\gamma} I^2/I^2_c)/(2\pi I_c) $,
where $\Phi_0$ is the magnetic flux quantum and $I_c$ is the Josephson
critical current. The nonlinear term originates from the
term $\propto \varphi^3$ in the Taylor expansion of the Josephson
supercurrent $I_J = I_c \sin\varphi$.
Generally, by exploiting the centrosymmetric nonlinearity of the supercurrent,
$I_J(-\varphi) = -I_J(\varphi)$
or, equivalently, the symmetric nonlinearity of inductance, $L_J(-I)=L_J(I)$,
JTWPAs can, however, only operate in the four-wave-mixing mode \cite{Agrawal}; i.e.,
when the signal ($\omega_s$), idler ($\omega_i$) and pump ($\omega_p$) tones
obey the relation $\omega_s + \omega_i = 2 \omega_p$.

\begin{figure}[b]
\begin{center}
\includegraphics[width=3.0in]{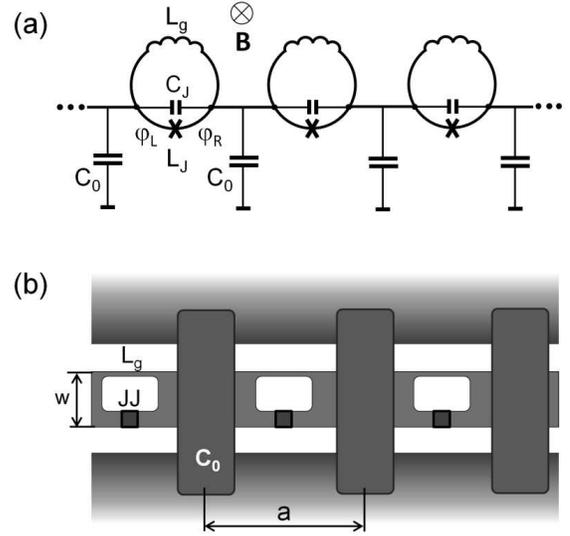}
\caption{(a) Electric diagram of the transmission line including an array of
one-junction SQUIDs. The tunnel JJ is presented as the parallel
connection of Josephson inductance $L_J$ and tunnel capacitance $C_J$.
(b) Possible layout of the transmission line with parallel plate capacitor
cross-overs.} \label{EqvSchm}
\end{center}
\end{figure}

In this paper, we propose a JTWPA possessing nonzero
quadratic nonlinearity produced
by nonlinear inductance $L_J(I)$, the power expansion of which contains a term proportional
to $I/I_c$ (or index $n$ having dependence on $\textbf{\textit{E}}$).
We engineer this nonlinearity by means
of modification of a current-phase relation in a superconducting circuit including JJs.
This property enables operation in a favorable, three-wave-mixing mode (whose
theory had been developed by Cullen as early as 1960 \cite{Cullen1960})
with frequencies obeying the relation
\begin{equation}
\omega_s + \omega_i = \omega_p.\label{3wave-mode}
\end{equation}
For such a regime the pump frequency shifts away substantially
from the signal band and therefore can be efficiently
filtered out from the amplified signal.
Because of inherently stronger quadratic interaction (in comparison to higher-order
cubic interaction), efficient operation in the three-wave-mixing mode
is possible with a smaller pump power. Moreover, the three-wave-mixing mode
enables a high dynamic range and wide-band operation. These promising characteristics
have been recently demonstrated in the experiment with a NbTiN kinetic-inductance
traveling-wave amplifier \cite{Vissers2016} in which quadratic nonlinearity
was created in addition to conventional Kerr nonlinearity of a thin superconducting
wire by means of applying a dc current bias.

\section{\label{sec:Model}The model}

The proposed ladder-type transmission line having \emph{full control} of both quadratic
and cubic nonlinearities consists of a serial array
of one-junction SQUIDs, or the so-called rf-SQUIDs, embedded
in the central conductor of the coplanar waveguide as shown in Fig.~1.
The value of the screening
SQUID parameter $\beta_L \equiv 2\pi L_g I_c/\Phi_0 < 1$  \cite{Braginski-Clark},
where $L_g$ is the geometrical inductance.
In this case, the external magnetic flux $\Phi_e$ induces
a flux inside the loop $\Phi_{\textrm{dc}}(\Phi_e)$ which is a single-valued
function of $\Phi_e$, and its value is found by
solving the transcendental equation (see, for example, Ref.~\cite{KK-book})
\begin{equation}
\Phi_{\textrm{dc}} + (\Phi_0/2\pi)\beta_L
\sin(2\pi\Phi_{\textrm{dc}}/\Phi_0) = \Phi_e.\label{Phi-vs-Phie}
\end{equation}
The phase drop across the JJ (see Fig. 1a) is therefore
$\varphi_{\textrm{dc}} \equiv \varphi_\textrm{L} - \varphi_\textrm{R} = 2\pi\Phi_{\textrm{dc}}/\Phi_0.$
The inverse (linear) inductance for a small input current is
$L^{-1}=L_g^{-1} + L_J^{-1}=L_g^{-1}(1+\beta_L \cos\varphi_{\textrm{dc}})$,
whereas the current-phase relation for the flux-biased SQUID is expressed by the formula
\begin{eqnarray}
I(\varphi) &=&  I_c \varphi /\beta_L + I_c [\sin (\varphi_{\textrm{dc}}+ \varphi) - \sin\varphi_{\textrm{dc}}] 
.\label{I-Phi}
\end{eqnarray}

\begin{figure}[b]
\begin{center}
\includegraphics[width=3.4in]{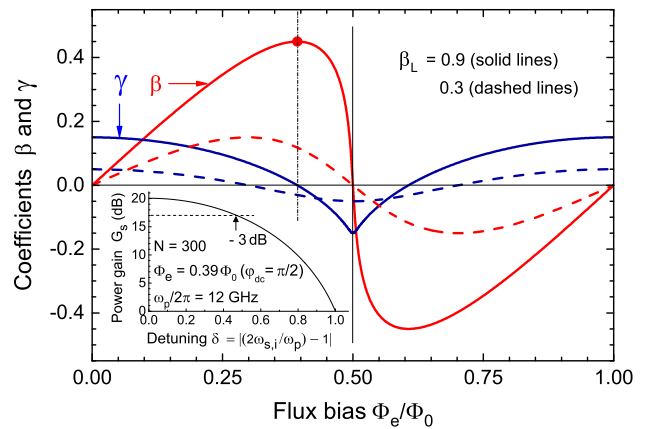}
\caption{Coefficients $\beta$ (red curves) and $\gamma$ (blue curves)
as functions of applied magnetic flux $\Phi_e$ for two values of SQUID
parameter $\beta_L$. The thin dotted vertical line indicates optimal flux bias $\Phi_e = 0.39 \Phi_0$
ensuring the phase $\varphi_{\textrm{dc}} = \pi/2$ corresponding
to maximum gain (maximum value of $|\beta|$ and zero value of $\gamma$) for the case of $\beta_L = 0.9$.
Inset shows the frequency dependence of signal-power gain $G_s$ in the JTWPA, calculated for
typical experimental parameters and the
optimum flux bias.}\label{beta-and-gamma}
\end{center}
\end{figure}

Here, $\varphi$ is the variation of the JJ phase associated with the (ac) current injected in the SQUID.
By expending the current $I(\varphi)$ in a power series of the small parameter $|\varphi|\ll 1$, one
arrives at the formula
\begin{equation}
I /I_c = (\beta_L^{-1} + \cos\varphi_{\textrm{dc}})\varphi - \tilde{\beta}\varphi^2 - \tilde{\gamma}\varphi^3 -...,\label{I-Phi-expd}
\end{equation}
with $\tilde{\beta}= \frac{1}{2} \sin\varphi_{\textrm{dc}}$,
$\tilde{\gamma} = \frac{1}{6} \cos\varphi_{\textrm{dc}}$, etc. The first term on the right-hand side
is related to the inverse linear
inductance of the SQUID, whereas $\tilde{\beta}$
and $\tilde{\gamma}$ describe the quadratic and Kerr (cubic) nonlinearities, respectively.
The quadratic nonlinearity introduces desired asymmetry $I(-\varphi)\neq -I(\varphi)$
and allows a number of remarkable physical effects
inaccessible with only Kerr nonlinearity. These include, for example,
the second harmonic generation (SHG) \cite{Franken1961}, spontaneous parametric down
conversion (SPDC) \cite{Klyshko1967,Burnham1970,Klyshko1988},
and what we focus on here, the three-wave mixing that enables parametric gain \cite{Cullen1960}.

Following the method of deriving a wave equation
for a ladder-type $LC$ transmission line with embedded JJs, which was
described in detail by Yaakobi et al. \cite{Yaakobi2013},
we arrive in our case at the equation
for the phase $\phi(x,t)$ on the circuit nodes,
\begin{eqnarray}
\frac{\partial ^2 \phi}{\partial x^2}-\omega^{-2}_0 \frac{\partial ^2 \phi}{\partial t^2}
+\omega^{-2}_J \frac{\partial ^4 \phi}{\partial x^2 \partial t^2}~~~~~~~~~~~~~~~~~~~~~~~ \nonumber\\
+\beta \frac{\partial}{\partial x} \left[ \left(\frac{\partial \phi}{\partial x}\right)^2 \right]
+\gamma \frac{\partial}{\partial x} \left[ \left(\frac{\partial \phi}{\partial x}\right)^3 \right]=0.
\label{wave-eq}
\end{eqnarray}
Here $x=X/a$ is a dimensionless coordinate; $a$, the section
size; $\omega_0=(LC_0)^{-\frac{1}{2}}$, the cutoff frequency; and,
$\omega_J=(LC_J)^{-\frac{1}{2}}$, the plasma frequency of the SQUID.
The ac part of the phase on a SQUID is $\varphi = a \frac{\partial \phi}{\partial X} = \frac{\partial \phi}{\partial x}$,
whereas voltage on a ground capacitor is
$V = (\Phi_0/2\pi)\frac{\partial \phi}{\partial t}$.
The nonlinear coefficients are
$\beta = \tilde{\beta}\beta_L$ and $\gamma =\tilde{\gamma} \beta_L$
(see their plots in Fig.~2).

The third term in Eq.~(\ref{wave-eq}) yields an ordinary superlinear frequency
dependence \cite{Yaakobi2013,Bell-Samolov2015} of the dimensionless wave vector $k=2\pi a/\lambda$,
\begin{equation}
k(\omega) =  \frac{\omega}{\omega_0\sqrt{1-\omega^2/\omega_J^2}},
\label{disp}
\end{equation}
which follows from a plane wave solution $e^{i(kx-\omega t)}$ of the corresponding
linear equation (at $\beta=\gamma=0$).
The section size $a$ is assumed to be much smaller than the wavelengths of
propagating signals, i.e. $a\ll \lambda$, the values of $k \ll 1$ and,
therefore, all working frequencies are small, $\omega_{s,i,p} \ll \omega_0$.
For sufficiently large plasma frequency $\omega_J \gg \omega_{s,i,p}$, the
chromatic dispersion Eq.~(\ref{disp}) is small, $k(\omega) \approx (\omega/\omega_0)(1+0.5\omega^2/\omega_J^2)$,
and, as usual in optics \cite{Agrawal},
positive, $dn(\omega)/d\omega >0$.

\section{\label{sec:Results}Analysis and characteristics}

\subsection{\label{sec:Results1}Signal gain}

The solution of the wave equation (\ref{wave-eq}) is found using the
coupled-mode-equation (CME) method \cite{Agrawal} in the form
\begin{equation}
\phi(x,t) = \frac{1}{2} \sum\limits_{j=\{s,i,p\}}{\left[A_j(x) e^{i(k_jx-\omega_j t)}+\textrm{c.c.}\right]},
\label{solution}
\end{equation}
where $k_{s,i,p}$ are wave vectors and $A_{s,i,p}(x)$ are slowly varying (i.e.
$\left| \frac{\partial^2A_j}{\partial x^2} \right| \ll k_j \left| \frac{\partial A_j}{\partial x} \right|
\ll k^2_j \left| A_j \right| $, $j = \{s, i, p\}$) amplitudes of the signal,
idler and pump waves, respectively.
Corresponding CMEs take the form
\begin{eqnarray}
\frac{dA_p}{dx} &=& i \frac{3}{8}\gamma k^3_p A_p |A_p|^2 -\frac{\beta}{2} k_s k_i A_s A_i e^{-i\Delta k x}, \label{coupled-mode-eq1} \\
\frac{dA_{s,i}}{dx} &=& i \frac{3}{4}\gamma k_{s,i} k^2_p A_{s,i} |A_p|^2 +\frac{\beta}{2} k_{i,s} k_p A^*_{i,s} A_{p} e^{i\Delta k x}.~~ \label{coupled-mode-eq2}
\end{eqnarray}
The phase mismatch resulted from dispersion equation (\ref{disp}) is
\begin{equation}
\Delta k = k_p - k_s - k_i \approx
\frac{\omega^3_p-\omega^3_s-\omega^3_i}{2\omega_0\omega^{2}_J}
= \frac{3\omega_s\omega_i\omega_p}{2\omega_0\omega^{2}_J} > 0,
\end{equation}
where the higher-order small terms $\propto \omega^{-1}_0
\omega^{-4}_J \omega_{s,i,p}^5$ are neglected. The terms
proportional to $\gamma$ produce
the self-phase modulation (SPM) $\vartheta_p$ and the cross-phase modulations (XPM) $\vartheta_s$ and $\vartheta_i$,
respectively \cite{Agrawal}, i.e.,
\begin{eqnarray}
\vartheta_p &=& (3/8) \gamma k^2_p |A_{p}|^2k_p, \label{theta-p-i-s1} \\
\vartheta_{s,i} &=& (3/4) \gamma k^2_p |A_{p}|^2 k_{s,i}, \label{theta-p-i-s2}
\end{eqnarray}
which contribute to the total
phase mismatch
\begin{equation}
\psi = \Delta k + \vartheta,
\end{equation}
where
\begin{equation}
\vartheta = \vartheta_p - \vartheta_s - \vartheta_i  \approx -(3/8)(\omega_p /\omega_0)^3 \gamma |A_{p}|^2.
\label{mismatch}
\end{equation}
Here we had again neglected the higher-order terms $\propto
(\omega_{s,i,p}/\omega_J)^2 (\omega_p /\omega_0)^3 \gamma$ because of a small $\gamma$
and relations $k_{s,i,p}\approx\omega_{s,i,p}/\omega_0$.

Under the undepleted-pump assumption, $|A_p(x)| = A_{p0} \gg |A_{s,i}(x)|$
the pump can be found from Eq.~(\ref{coupled-mode-eq1}) explicitly,
\begin{equation}
A_p(x) = A_{p0} e^{i \chi_{p0}}e^{i \vartheta_p x},
\end{equation}
where $\chi_{p0}$ is the initial phase. Then Eq.~(\ref{coupled-mode-eq2})
is simplified:
\begin{eqnarray}
\frac{dA_{s,i}}{dx} &=& i\vartheta_{s,i} A_{s,i} + 2 g_0 \frac{\omega_{i,s}}{\omega_p}
A^*_{i,s} e^{i \chi_{p0}}e^{i \vartheta_p x} e^{i \Delta k x}, \label{coupled-mode-As}  \\
g_0 &=& |\beta| A_{p0} \omega^2_p/4\omega^2_0.  \label{g0-expression}
\end{eqnarray}
For the zero initial idler, $A_i(0) = 0$, but
nonzero initial signal, $A_{s}(0) = A_{s0} e^{i\chi_{s0}} \neq 0$, the solution of
linear equations (\ref{coupled-mode-As})
can be presented in the form \cite{Agrawal}
\begin{eqnarray}
A_{s}(x)= A_{s0} e^{i\chi_{s0}} \left[ \cosh(g x) - \frac{i\psi}{2g}\sinh(g x) \right] e^{i(\vartheta_s +\frac{\psi}{2})x},~ \\
A_{i}(x)=2\frac{g_0 \omega_s}{g \omega_p} A_{s0} e^{i\chi_{i0}} e^{-i\chi_{p0}} \sinh(g x) e^{i(\vartheta_i + \frac{\psi}{2})x}~~~~~~~~~~
\label{As-Ai}
\end{eqnarray}
with initial phases obeying the relation $\chi_{p0}-\chi_{s0}-\chi_{i0} = 0$. The complex exponential gain factor is
\begin{equation}
g =  \left[(1-\delta^2)g^2_0 -(\psi/2)^2\right]^{\frac{1}{2}},
\label{gain-factor}
\end{equation}
where $1-\delta^2 $ stands for $4\omega_s\omega_i/\omega^2_p$.
Here the dimensionless detuning is
\begin{equation}
\delta = |2\omega_{s,i}- \omega_p|/\omega_p,
\end{equation}
while the total phase mismatch equals
\begin{equation}
\psi = \Delta k + \vartheta= \frac{3\omega_p^3}{8\omega_0^3}
\left[ (1-\delta^2) \frac{\omega^2_0}{\omega^2_J} - \gamma A^2_{p0} \right]. \label{mismatch-psi}
\end{equation}
For the zero phase mismatch $\psi=0$, the signal has the maximum power gain
\begin{equation}
G_s =|A_{s}/A_{s0}|^2 = 1 + \sinh^2 \left(\sqrt{1-\delta^2}g_0 N \right),
\label{gain-Gs}
\end{equation}
where $N$ is the length of the line.

\subsection{\label{sec:Results2}Phase mismatch}

Generally, realization of zero phase mismatch is a
challenging problem in designing \emph{four-wave-mixing} JTWPAs.
The main difficulty is that both the phase modulation and the gain
depend on the very same parameter, i.e. the Kerr nonlinearity $\gamma$.
This problem can be partially solved by very neat engineering of dispersion
using auxiliary resonance elements embedded in the transmission line \cite{OBrien2014,White2015}.
In contrast to this situation, in our \emph{three-wave-mixing} case the phase-modulation effects and signal gain
are controlled by \emph{two different} nonlinear terms, i.e. $\propto \gamma A^2_{p0}$
and $\propto \beta A_{p0}$, respectively, so a phase matching can be realized
by simply adjusting external flux bias $\Phi_e$ (see the
dependencies $\beta(\Phi_e)$ and $\gamma(\Phi_e)$ in Fig.~2).

In particular, for $\omega_s \approx \omega_i \approx 0.5 \omega_p$ (i.e. small $\delta \approx 0$)
the total phase mismatch $\psi$ Eq.~(\ref{mismatch-psi}) can be made
equal to zero if, first, $\gamma >0$
and, second, $\gamma A^2_{p0}$ is of the order of $\omega^2_0/\omega^2_J = C_J/C_0$.
Assuming that the optimal pump amplitude is determined
by a swing of the Josephson phase $\varphi_a \sim 1$, i.e. $A_{p0} = \varphi_a/k_p \approx \omega_0/\omega_p$, we
obtain $\gamma \approx \omega^2_p/\omega^2_J \ll 1$. In this case, finite detuning
$\delta$ may cause phase mismatch $|\psi| \approx (3/8)\delta^2 \omega^3_p /(\omega_0 \omega^2_J)$
which is much smaller than the exponential gain
factor $g = 0.5 \sqrt{1-\delta^2} |\beta| \omega_p/\omega_0 $
(of course, not for  vanishingly small values of $|\beta|$ and a value $\delta$ not very close to 1).
Thus, for proper circuit parameters, an exponential
gain with frequency dependence given by Eq.~(\ref{gain-Gs}) is always possible.

Large exponential power gain allows operating even in the regime of
imperfect phase matching, $\psi \neq 0$. The dephasing length
on which the phase mismatch becomes substantial is
\begin{equation}
N_{\psi} = \frac{\pi}{|\psi|}
= \frac{8\pi \omega^3_0}{3\omega^3_p}
\left| \frac{(1-\delta^2)\omega^2_0}{\omega^2_J} - \gamma A^2_{p0}\right|^{-1}. \label{dephasing}
\end{equation}
This formula can be rewritten as
\begin{equation}
N_{\psi} = \frac{N_{\textrm{D}}N_{\textrm{PM}}}{| N_{\textrm{D}} - N_{\textrm{PM}}|} = | N_{\textrm{D}}^{-1} - N_{\textrm{PM}}^{-1}|^{-1}, \label{dephasing-length}
\end{equation}
where
\begin{equation}
N_{\textrm{D}} = \frac{8\pi \omega_0\omega^2_J}{3\omega^3_p(1-\delta^2)} \label{Ndephasing1}
\end{equation}
and
\begin{equation}
N_{\textrm{PM}}=\frac{8\pi \omega^3_0}{3\gamma\omega^3_p A^2_{p0}} \label{Ndephasing2}
\end{equation}
are the dephasing lengths attributed solely to chromatic dispersion and
attributed solely to SPM and XPM, respectively.
For optimal values of $|\beta| = \beta_L/2$ and $\gamma \rightarrow 0$ (see Fig.~2)
the length $N$ which is sufficient for attaining designed value
of gain $G_s$,
\begin{equation}
N = \frac{\textrm{arccosh}\sqrt{G_s}}{g} \approx \frac{4 \omega^2_0 \ln(2\sqrt{G_s})}{|\beta|\omega^2_p(1-\delta^2)A_{p0}} , \label{Ndephasing}
\end{equation}
can safely be designed substantially smaller than very large $N_{\psi}$.
In this case the gain suppression because of the pump depletion may dominate.

\subsection{\label{sec:Results3}Pump depletion}

Depletion of the pump power in this JTWPA may occur by means of
two major mechanisms. Firstly, there is a possible leak of power to higher harmonics,
especially to the second harmonic $2\omega_p$.
An analysis of the corresponding CME shows
that because of weak frequency dispersion ($2\omega_p \ll \omega_J$) and both small SPM and XPM
($\gamma \ll 1$) the phase mismatch between the main tone and the second harmonics is small.
Without special precautions this fact may cause intensive SHG.
To suppress SHG one may create a narrow stopband around $2\omega_p$ by applying,
for example, the technique of periodic variation of wave impedance developed
by Eom et al. \cite{Eom2012}.
In our case, such stopband engineering is reduced to some change of, for
example, capacitance $C_0$ in every
$m-$th section of the line. The number $m$ should correspond to
a half wavelength of a preselected frequency, i.e., $m = [2\pi \omega_0/4\omega_p]$.
Alternatively, one can reduce the cutoff frequency $\omega_0$ to prevent propagation
of higher ($\geq 2$) pump harmonics having frequencies larger than $\omega_0$,
as was done in Ref.~\cite{White2015}.

Secondly, the input pump power $P_{p0}$ is inevitably converted into the power
of signal $P_s$ and idler $P_i$ ($P_{j} \propto (\frac{\partial \phi_j}{\partial t})^2
 = \omega^2_j |A_j|^2$, $j = \{s,i,p\}$), i.e.
\begin{equation}
\omega^2_p A_{p0}^2 + \omega^2_s A_{s0}^2 = \omega^2_p |A_p|^2 + \omega^2_s |A_s|^2 + \omega^2_i |A_i|^2. \label{power balance}
\end{equation}
Specifically, for a zero phase mismatch, the power gain versus the length dependence
is described by the formula (see the Appendix for details)
\begin{equation}
G^{\textmd{(d)}}_s = \frac{1}{\textrm{dn}^2( g N/\textmd{k},\textmd{k})} \rightarrow 1+\sinh^2(g N),~~\textrm{as}~~ \textmd{k}\rightarrow 1, \label{A_s-depleted}
\end{equation}
where $\textrm{dn}(u,\textmd{k})$ is the Jacobi elliptic
function with modulus value
\begin{equation}
\textmd{k}= \sqrt{\omega_p A^2_{p0}/(\omega_s A^2_{s0}+\omega_p A^2_{p0})}. \label{mod-k-value}
\end{equation}
The maximum possible power gain $G^{\textrm{max}}_s = 1/(1-\textmd{k}^2)$ is
achieved for the line length $N = \textmd{k} K(\textmd{k})/g$,
where $K$ is the complete elliptic integral of the first kind.
For a longer line the gain is reduced due to reverse power
conversion from the signal and the idler to the pump \cite{Cullen1960}.

\section{\label{sec:Exp}Possible design and discussion}

Finally, let us make an estimation of possible experimental parameters.
For state-of-the-art four-wave-mixing counterparts critical
current $I_c = 5~\mu$A \cite{White2015,Macklin2015}. Assuming a similar value of $I_c$ and
a critical-current density of 500~A/cm$^2$, the tunnel area is about 1~$\mu \textrm{m}^2$
and  JJ capacitance $C_J \approx$ 60~fF. For a width of center strip of $w = 15~\mu$m
capacitance $C_0$
can be designed \cite{Mutus2014} to be around 100~fF without
significant increase of the section size $a \sim 2w$. By applying small meandering
geometrical inductance
$L_g$ can be made to be about 57~pH, yielding $\beta_L \approx 0.9$.
For appropriate flux bias
corresponding to $\varphi_{\textrm{dc}} = \pi/2$
and ensuring maximum value of $\beta = \beta_L/2 = 0.45$ and zero $\gamma$ (see Eq.~(\ref{Phi-vs-Phie})),
the resulting inductance $L = L_g/(1+\cos \varphi_{\textrm{dc}}) = L_g = 57$~pH.

The wave impedance of this line is, therefore,
$Z_0 = (L/C_0)^{\frac{1}{2}} \approx 24~\Omega$, plasma frequency
$\omega_J/2\pi \approx 86$~GHz and the cutoff frequency $\omega_0/2\pi \approx 67$~GHz.
Taking the pump frequency $\omega_p/2\pi = 12~\textrm{GHz}$ and
amplitude $A_{p0} = 0.5 \omega_0/\omega_p \approx 2.8$,
corresponding to phase swing $\varphi_a = k_p A_{p0} = 0.5~\textrm{rad} \lesssim 30^\circ$
and current $I^{\textrm{rms}}_{p0} = \frac{1}{\sqrt{2}}(\Phi_0/2\pi L)\varphi_{a}
\approx 1.97~\mu\textrm{A}$ ($ \approx -70.3$~dBm),
we obtain the factor $g_0 = |\beta| \omega_p/8\omega_0 \approx 0.01$ (see Eq.~(\ref{g0-expression})).
The power gain of $N$-section JTWPA given by Eq.~(\ref{gain-Gs}) is, therefore,
$G_s \approx 1 + \sinh^2(0.01 N)$, i.e. about 20~dB for $N = 300$,
whereas the geometrical length $aN \approx$ 9~mm. This length corresponds
to $n_\lambda =aN/\lambda_p= N/(2\pi\omega_0/\omega_p) \approx 8.6$ wavelengths
of the pump.

The evaluated dephasing lengths in Eqs.(\ref{Ndephasing1}) and (\ref{Ndephasing2})
are $N_{\textrm{D}}= 2400$ and $N_{\textrm{PM}}
186/\gamma$, respectively. Taking the most unfavorable value of $\gamma = -0.1$
(achieved instantly only at the extreme
position of an oscillating phase $\varphi$), we have a very conservative
estimate Eq.(\ref{dephasing-length}) for a total dephasing length $N_\psi  \approx 1000 \gg N$.
The dependence of the power gain $G_s$ on frequency (shown in the inset of Fig.~2) yields
a remarkably wide 3 dB range of $ 0.47\omega_p/2\pi \approx$ 5.64~GHz.
While this 3 dB range is only slightly larger than that achieved in the four-wave
mixing JTWPA (e.g., $\approx 0.42\omega_p/\pi$ \cite{Bell-Samolov2015}, $0.38 \omega_p/\pi$ \cite{White2015}
and $0.21\omega_p/\pi$ \cite{Macklin2015}), the latter
typically exhibits considerable ripple or other variations in the gain
versus frequency, which can be practically cumbersome.
Moreover, the contiguous frequency range of the power gain in our JTWPA is
extended up to a pump frequency of $\omega_p/2\pi =$ 12~GHz.

\begin{figure}
\begin{center}
\includegraphics[width=3.4in]{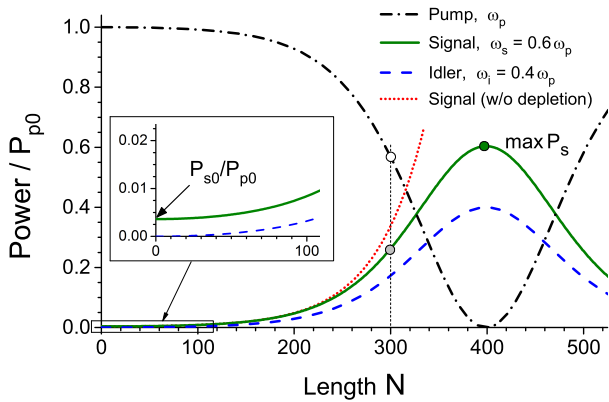}
\caption{Amplification of signal (green solid curve) and idler (dashed blue curve),
and depletion of pump (black dash-dotted curve)
versus the line length $N$ found by numeric solving of CMEs.~(A1)-(A3)
in the case of $\omega_s = 0.6 \omega_p$.
The gray and open circles show the output powers of the signal and the pump, respectively,
corresponding to a 1 dB suppression of the nominal gain of 20 dB in the 300-section array
occurred at $\textmd{k}^{\prime} = 0.075$.
For the pump power $P_{p0} = -70.3$~dBm the signal-saturation power is
$P_{s0} = (\textmd{k}^{\prime}/\textmd{k})^2 (\omega_s/\omega_p) P_{p0}
\approx 0.0057 (\omega_s/\omega_p) P_{p0} \approx -96$~dBm.
For comparison, red dotted curve shows the exponential rise (described by Eq.~(\ref{gain-Gs}))
of this signal in the case of undepleted pump.
The solid green circle shows maximum signal power
$P_s = (\omega_s/\omega_p)P_{p0} \approx -73.5$~dBm corresponding to a maximum
achievable gain of about 22.5 dB
(at $N = N_m = 399$).} \label{Depl}
\end{center}
\end{figure}

As follows from the corresponding coupled equations (A1)-(A3) (see the Appendix) the 1~dB
reduction of the gain caused by pump depletion occurs when the complementary modulus
value $\textmd{k}^{\prime} \equiv \sqrt{1-\textmd{k}^2} \approx 0.075$.
The simulated dependence of the signal and pump powers versus the length $N$ presented in Fig.~3
shows that, for a rather modest pump power of $-70.3$~dBm, the input signal-saturation
power is $P_{s0} \approx - 96$~dBm (cf. $-98$~dBm for a 20-dB gain in Refs. \cite{OBrien2014,Bell-Samolov2015}
and $-92$~dBm for a 12-dB gain in Ref. \cite{White2015}).
The dynamic range can be further improved by an increase of the phase swing $\varphi_a$ from
0.5~rad up to, for example, 1~rad with a corresponding 6-dB increase of both the pump and the
signal-saturation powers,
i.e., up to approximately $-64.3$~dBm and $-90$~dBm, respectively.

For the above-mentioned experimental parameters, the
line impedance $Z_0$, is still below 50~$\Omega$ which may require special
matching with a signal source and a post-amplifier. This matching can be done, for example,
with the help of a Klopfenstein taper \cite{Klopfenstein1956} having a design similat to that
in Ref.~\cite{Mutus2014}.
The low-$Z_0$ problem can be solved (at a price of some reduction of dynamic range
and larger size of the circuit)
by simultaneously making ground capacitance $C_0$ and the critical current $I_c$ smaller and
the geometrical inductance $L_g$ larger.
For example, $I_c = 2.5~\mu$A, $L_g =114$~pH
(which can be realized by means of a relatively large size inductor)
and $C_0 =50$~fF should yield $Z_0 \approx 50$~$\Omega$, while keeping $\beta_L$,
$\omega_J$ and $\omega_0$ equal to the above-mentioned design values.
To keep the length of such circuit reasonably
short one can, for example, replace large geometrical inductance $L_g \propto 1/I_c$ by
kinetic inductance of a serial array of two to four larger JJs.
Finally, another strategy for increasing $Z_0$ toward 50~$\Omega$ is replacing
a single rf-SQUID in each section by a group of two to four serially connected rf-SQUIDs.

\section{Conclusion}

We have developed a concept of a JTWPA with three-wave mixing,
which can outperform state-of-the-art JTWPAs
operating on the principle of four-wave mixing. Our circuit enables high
gain, the widest flat bandwidth and nominally zero phase matching. Moreover,
this JTWPA allows efficient operation with slightly imperfect phase matching.
The proposed design is simple, compact, excludes the engineering
of sophisticated resonant phase-matching elements, allows cascading and multiplexing,
and is feasible in the labs with standard fabrication facilities.

This JTWPA flexibly allows further optimization of its
parameters and possible integration with, for example, SQUID transducers,
single-photon detectors, qubits, etc. We believe that this amplifier
with potentially quantum-limited noise performance will advance high-fidelity
measurements and signal processing at the single-photon level.
Last, but not least, realization of a SPDC regime \cite{Klyshko1988}
in a circuit with a Josephson noncentrosymmetric nonlinear medium,
like in this JTWPA, may allow its application for the generation of
entangled photon pairs with frequencies obeying the relation
$\omega_1 + \omega_2 = \omega_p$. Such entangled biphotons are of particular
importance for creating a quantum processor, quantum-key distribution, and a
secure transmission of data.

\section{Acknowledgments}

The author thanks R. Dolata, M. Khabipov, G. Kh. Kitaeva, M. Chekhova, P. Meeson, K. Porsch, P. Delsing,
and J. Martinis for stimulating discussions. This work belongs partly to the Joint Research Project MICROPHOTON 
of the European Metrology Research Programme (EMRP). The EMRP is jointly funded by the EMRP participating 
countries within EURAMET and the European Union.

\section{\label{sec:App}Appendix: Effect of pump depletion}

\renewcommand{\theequation}{A\arabic{equation}}
\setcounter{equation}{0}  

Neglecting the SPM and XPM effects (i.e. $\gamma \rightarrow 0$) and the phase mismatch
due to chromatic dispersion ($\Delta k \rightarrow 0$) the coupled-mode equations (8)
and (9) for the amplitudes and the phases of the waves,
$A_{p,s,i}(x) = |A_{p,s,i}(x)| e^{i\chi_{p,s,i}(x)}$, take the form:
\begin{eqnarray}
\frac{d|A_p|}{dx} &=& - \frac{\beta}{2} \frac{\omega_s \omega_i}{\omega^2_0} |A_s| |A_i| \cos \chi, \label{amp-coupled-mode-eq1} \\
\frac{d|A_s|}{dx} &=& \frac{\beta}{2} \frac{\omega_i \omega_p}{\omega^2_0} |A_i| |A_{p}| \cos \chi, \label{amp-coupled-mode-eq2} \\
\frac{d|A_i|}{dx} &=& \frac{\beta}{2} \frac{\omega_s \omega_p}{\omega^2_0} |A_s| |A_{p}| \cos \chi, \label{amp-coupled-mode-eq3} \\
\frac{d\chi}{dx} &=& \frac{\beta }{2} \frac{\omega_p \omega_s \omega_i}{ \omega^2_0} \nonumber\\
&\times& \left( \frac{|A_s| |A_i|}{\omega_p|A_{p}|} - \frac{|A_p| |A_i|}{\omega_s|A_{s}|} - \frac{|A_s| |A_p|}{\omega_i|A_{i}|} \right) \sin \chi,~~~~~~
\end{eqnarray}
where $\chi = \chi_p - \chi_s - \chi_i$.
The initial conditions
read $|A_p(0)| = A_{p0}$, $|A_s(0)| = A_{s0}$, $|A_i(0)| = 0$ and
$\chi(0) = \chi_{p0} - \chi_{s0} - \chi_{i0} = 0$.
Therefore, the fourth equation (A4) yields constant phase $\chi = 0$, so
$\cos \chi$ is equal to 1.
The set of equations (A1)-(A3) implies that
\begin{equation}
- \omega_p|A_p|\frac{d|A_p|}{dx} =  \omega_s|A_s|\frac{d|A_s|}{dx} = \omega_i|A_i|\frac{d|A_i|}{dx},
\end{equation}
or, equivalently, in terms of wave powers ($P_j \propto \omega^2_j |A_j|^2 $, $j = \{p,s,i\}$),
\begin{equation}
- \frac{1}{\omega_p}\frac{dP_p}{dx} =  \frac{1}{\omega_s}\frac{dP_s}{dx} = \frac{1}{\omega_i}\frac{dP_i}{dx}.
\end{equation}
Virtually, these are the Manley-Rowe relations for the waves \cite{Manley-Rowe1956}.
The three conserved quantities (in the sense that they are spatially invariant) are
\begin{equation}
\frac{P_p}{\omega_p} + \frac{P_s}{\omega_s} = M_1,~~~~\frac{P_p}{\omega_p} + \frac{P_i}{\omega_i} = M_2,~~~ \frac{P_s}{\omega_s} - \frac{P_i}{\omega_i} = M_3.
\end{equation}
Using the initial conditions, the constants $M_1$, $M_2$, and $M_3$  entering Eq.~(A7) can be found, so
the corresponding relations between the wave amplitudes read
\begin{eqnarray}
\omega_p |A_p|^2 + \omega_s |A_s|^2&=& \omega_p |A_{p0}|^2 + \omega_s |A_{s0}|^2, ~~~~~~~~\\
\omega_p |A_p|^2 + \omega_i |A_i|^2&=& \omega_p |A_{p0}|^2, \\
\omega_s |A_s|^2 - \omega_i |A_i|^2&=& \omega_s |A_{s0}|^2.
\end{eqnarray}
Substituting expressions
\begin{equation}
|A_s|= \sqrt{ |A_{s0}|^2 + (\omega_p/ \omega_s)(|A_{p0}|^2 - |A_p|^2)}
\end{equation}
and
\begin{equation}
|A_i|= \sqrt{(\omega_p/\omega_i)( |A_{p0}|^2- |A_p|^2)}~~~~~~~~~~~~~
\end{equation}
into Eq.~(A1),
\begin{equation}
|A_i|= \sqrt{(\omega_s/\omega_i)( |A_{s}|^2- |A_{s0}|^2)}~~~~~~~~~~~~~
\end{equation}
and
\begin{equation}
|A_p|= \sqrt{ |A_{p0}|^2 + (\omega_s/ \omega_p)(|A_{s0}|^2 - |A_s|^2)}
\end{equation}
into Eq.~(A2) and
\begin{equation}
|A_s|= \sqrt{ |A_{s0}|^2 + (\omega_i/ \omega_s)|A_i|^2}~~~~~~~~~~~~~~
\end{equation}
and
\begin{equation}
|A_p|= \sqrt{|A_{p0}|^2- (\omega_i/\omega_p) |A_i|^2}~~~~~~~~~~~~~~
\end{equation}
into Eq.~(A3) we obtain a set of three uncoupled equations which
are solved by separation of the variables,
\begin{eqnarray}
&&\frac{d|A_p|}{\sqrt{\frac{\omega_p}{\omega_i}( |A_{p0}|^2- |A_p|^2) [|A_{s0}|^2 + \frac{\omega_p}{\omega_s}(|A_{p0}|^2 - |A_p|^2)]}}  \nonumber \\
&&~~~~= -\frac{\beta}{2} \frac{\omega_s \omega_i}{\omega^2_0} dx,   \\
&&\frac{d|A_s|}{\sqrt{\frac{\omega_s}{\omega_i}( |A_{s}|^2- |A_{s0}|^2) [|A_{p0}|^2 + \frac{\omega_s}{\omega_p}(|A_{s0}|^2 - |A_s|^2)]}}  \nonumber \\
&&~~~~=\frac{\beta}{2} \frac{\omega_i \omega_p}{\omega^2_0} dx,   \\
&&\frac{d|A_i|}{\sqrt{[|A_{s0}|^2 + \frac{\omega_i}{\omega_s}|A_i|^2] [|A_{p0}|^2- \frac{\omega_i}{\omega_p}|A_i|^2]}}  \nonumber \\
&&~~~~=\frac{\beta}{2} \frac{\omega_s \omega_p}{\omega^2_0} dx.
\end{eqnarray}
The solutions (see, for example, Ref. \cite{Rudenko1977}) can be expressed in
terms of the Jacobi elliptic functions $\textrm{sn}(u, \textmd{k})$,
$ \textrm{cn}(u, \textmd{k})$, and $\textrm{dn}(u, \textmd{k})$, i.e.
\begin{eqnarray}
|A_p|&=& |A_{p0}| ~\textrm{sn}(K - gN/\textmd{k}, \textmd{k}), ~~~~~~~~~~~~~~~~~~~~~~ \\
|A_s|&=& \sqrt{\frac{\omega_p}{\omega_s}} |A_{p0}| ~\textrm{dn}(K - gN/\textmd{k}, \textmd{k}) \nonumber \\
  &=& |A_{s0}|/\textrm{dn}(gN/\textmd{k}, \textmd{k}),~~~~~ \\
|A_i|&=& \sqrt{\frac{\omega_p}{\omega_i}} |A_{p0}| ~\textrm{cn}(K - gN/\textmd{k}, \textmd{k}).
\end{eqnarray}
Here $K$ is the complete elliptic integral of the first kind with the
modulus value $\textmd{k}$ given by Eq.~(\ref{mod-k-value}).
The complementary modulus value is
\begin{equation}
\textmd{k}^{\prime} = \sqrt{1- \textmd{k}^2} = \sqrt{\omega_s A^2_{s0}/(\omega_s A^2_{s0}+\omega_p A^2_{p0})}
\end{equation}
and the ratio
\begin{equation}
\textmd{k}^{\prime}/\textmd{k} = \sqrt{\omega_s/\omega_p} A_{s0}/A_{p0},
\end{equation}
while $gN$ stands
for $\frac{\omega_p \sqrt{\omega_s \omega_i}}{\omega^2_0}\beta A_{p0} N = \sqrt{1-\delta^2}g_0N$
(see Eqs.~(\ref{g0-expression}) and (\ref{gain-factor}) in the main text). Formula Eq.~(\ref{A_s-depleted}) for power gain $G^{\textmd{(d)}}_s$
immediately follows from Eq.~(A21).

The transmission line length ensuring maximum possible power
gain
\begin{equation}
G^{\textrm{max}}_s =  \frac{1}{{\textmd{k}^{\prime}}^2} = \frac{\omega_s A^2_{s0}+\omega_p A^2_{p0}}{\omega_s A^2_{s0}} \label{gain-max}
\end{equation}
should correspond to the half of the period of the Jacobi elliptic function $\textrm{dn}(u, \textmd{k}) = \textrm{dn}(u + 2K, \textmd{k})$  \cite{Gradshteyn-and-Ryzhik},
i.e. $K$, so
\begin{equation}
N_{m} = \frac{\textmd{k}}{g}~K(\textmd{k}) \approx \frac{1}{g}\ln\frac{4}{\textmd{k}^{\prime}}.
\end{equation}
For a longer transmission line, in our case at $N > N_{m} = 399$, the reverse power
conversion from the signal and the idler to the pump occurs as shown in Fig.~3
(see also Fig.~6 in Ref.~\cite{Cullen1960}).

\end{document}